\documentclass[5p]{elsarticle}
\usepackage{graphicx, amsmath, amssymb, aas_macros, xspace}
\usepackage{dcolumn}
\usepackage{natbib, xfrac}
\usepackage{hyperref}
\usepackage{bm, color}
\graphicspath{{}{./fig/}{./png/}}

\newcommand{\Fig}[1]{Figure~\ref{#1}}
\newcommand{\Eq}[1]{Equation~(\ref{#1})}
\newcommand{\Tab}[1]{Table~\ref{#1}}
\newcommand{\EQ}{\begin{equation}}
\newcommand{\EN}{\end{equation}}

\newcommand{\cms}{\,cm\,s$^{-1}$\xspace}

\newcommand{\gcm}{\,g\,cm$^{-3}$\xspace}
\newcommand{\dcm}{\,dyn\,cm$^{-1}$\xspace}
\newcommand{\cc}{\emph{cc}\xspace}
\newcommand{\ccs}{{\emph{cc}}s\xspace}

\biboptions{authoryear}

\date{\today,~ $ $Revision: 1.2 $ $}
\begin{document}

\begin{frontmatter}

\title{Compound Chondrules fused Cold}

\author[AIH]{Alexander Hubbard\corref{cor1}}
\ead{ahubbard@amnh.org}


\cortext[cor1]{Corresponding author}

\address[AIH]{Department of Astrophysics, American Museum of Natural History, New York, NY 10024-5192, USA;
Phone: 212-313-7911}

\begin{abstract}
About $4-5\%$ of chondrules are compound: two separate chondrules stuck together. This is commonly believed to be
the result of the two component chondrules having collided shortly after forming, while still molten.
This allows high velocity impacts to result in sticking.
However, at $T\sim 1100$\,K,
the temperature below which chondrules collide as solids (and hence usually bounce),
coalescence times for droplets of appropriate composition are measured in tens of seconds.
Even at $1025$\,K, at which temperature theory predicts that the chondrules must have collided extremely slowly
to have stuck together,
the coalescence time scale is still less than an hour.
These coalescence time scales are too short for the collision of molten chondrules to explain the observed frequency
of compound chondrules.
We suggest instead
a scenario where chondrules stuck together in slow collisions while fully solid; and the resulting chondrule pair was subsequently briefly heated
to a temperature in the range of $900-1025$\,K.  In that temperature window the coalescence time is finite but long, covering a span of hours to a decade.
This is particularly interesting because those temperatures are  precisely the critical window for thermally
ionized MRI activity, so compound chondrules provide a possible probe into that vital regime.
\end{abstract}


\end{frontmatter}


\section{Introduction}

\begin{figure}[t!]\begin{center}
\includegraphics[width=0.75\columnwidth]{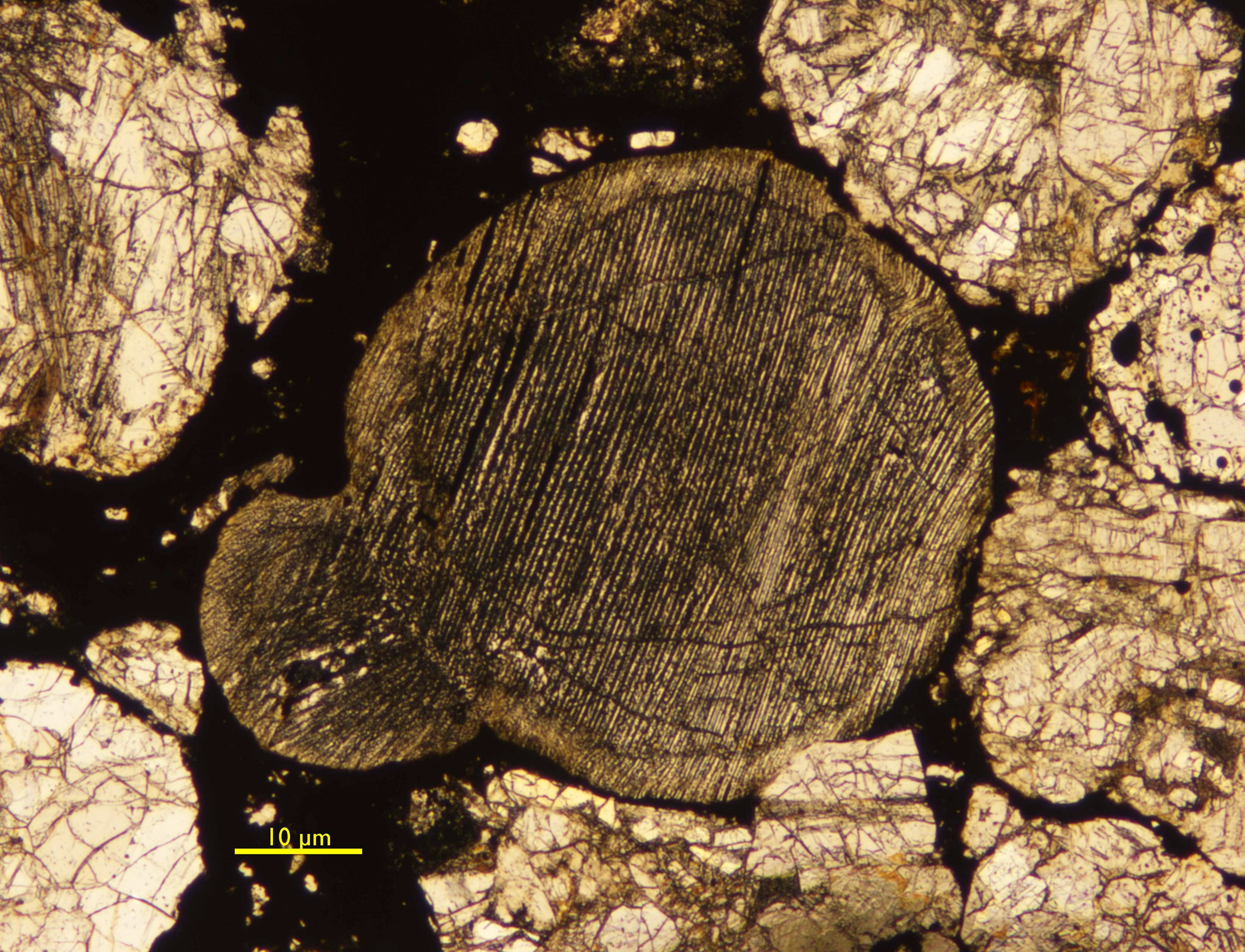}
\end{center}
\caption{
A plane polarized light micrograph of a compound chondrule from the Semarkona thin section AMNH 4128-1.
\label{fig_pic} 
}
\end{figure}

Chondrules are sub-millimeter sized igneous inclusions found in
chondritic meteorites which were melted in the Solar Nebula, the
gas and dust disk from which our Solar System formed.  Usually distinct objects within their host meteorite, they are
sometimes found as part of a compound chondrule: two distinguishable chondrules fused together (see \Fig{fig_pic}).
While compound chondrules (\ccs, lower-case to distinguish from carbonaceous chondrites) are rare, with a frequency of about $4-5\%$ of chondrules
\citep{1981Metic..16...17G,2004M&PS...39..531C}, they
are nonetheless common enough for some basic statistical information about their nature to be
well established \citep{1981Metic..16...17G,1995GeCoA..59.1847W}.  
\cite{1995GeCoA..59.1847W} distinguishes
 between sibling and independent \ccs,
depending on whether the primary and secondary chondrules either have or don't have similar composition and textural types.
One of the greatest difficulties in studying compound chondrules
is the difficulty in correcting 2D information from thin sections to 3D properties \citep{2004M&PS...39..531C},
which makes determining their precise frequency difficult, and establishing other
parameters, such as the size ratio of the secondary chondrule to the primary, fraught.

\begin{figure}[t!]\begin{center}
\includegraphics[width=0.75\columnwidth]{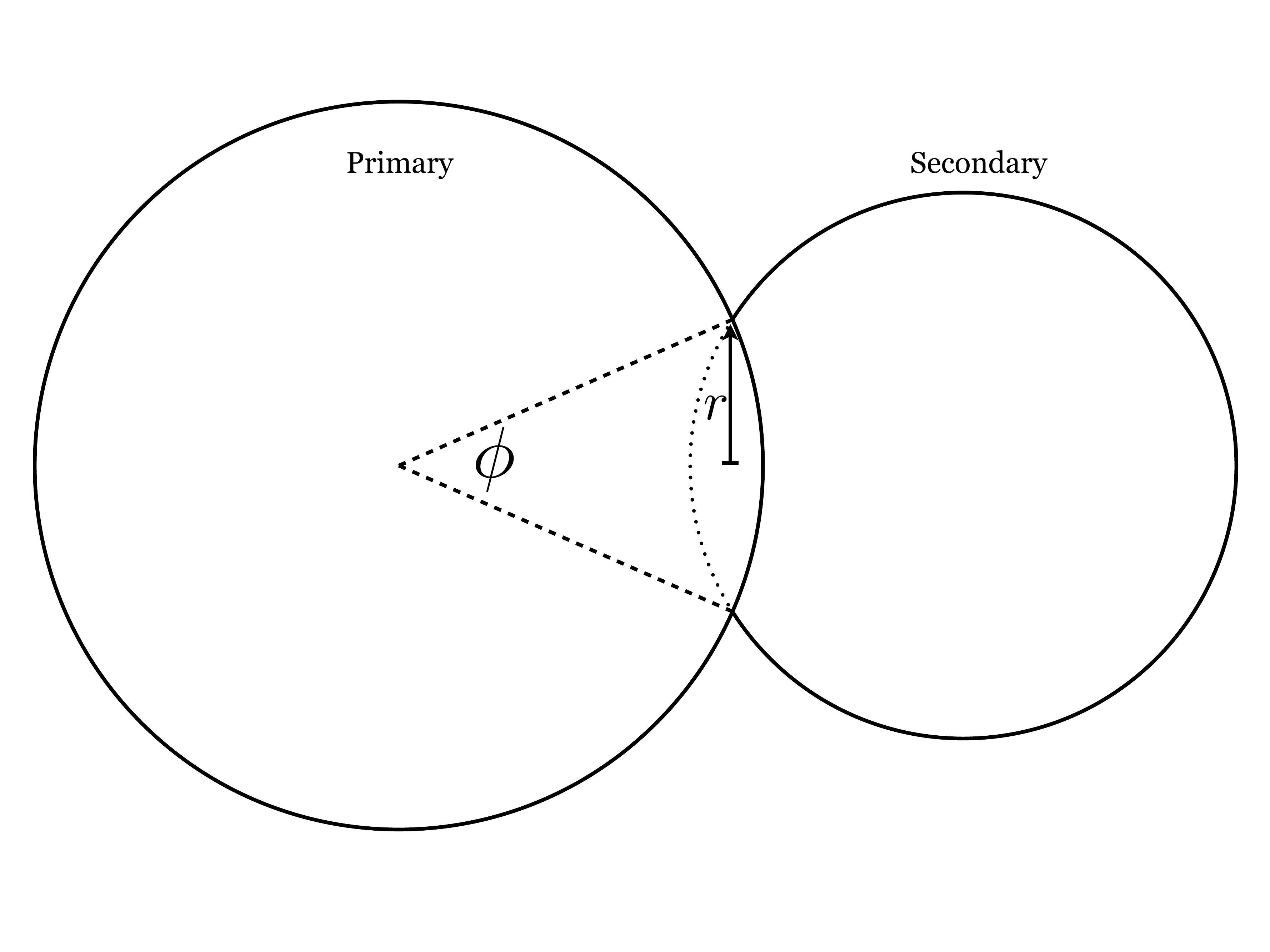}
\end{center}
\caption{
Cartoon of a compound chondrule.  The neck of the bridge has radius $r$, while 
$\phi$ is the contact angle.
\label{fig_cartoon} 
}
\end{figure}

\cite{1981Metic..16...17G} proposed that compound chondrules were made by collisions between still
molten chondrules shortly after they were formed from ambient dust.  If so, then the \cc  frequency 
is an important constraint because it links the chondrule-chondrule
relative velocity and cooling times to the chondrule number density \citep[e.g.][]{2012M&PS...47.1139D}.
However, sibling and independent \ccs occur at similar frequencies, implying that if the \cc fusing process
occurred at temperatures where the chondrules were molten, that
process must have been able to maintain the structural integrity of droplets of nearly identical liquids
in close contact with each other.  This requirement is exacerbated by the observation that small contact angle
\ccs dominate the statistics \citep{1995GeCoA..59.1847W,2004M&PS...39..531C},
i.e.~the radius of the neck between the two component chondrules is generally much smaller
than the radius of the individual chondrules, see \Fig{fig_cartoon}.



\cite{1995GeCoA..59.1847W} suggests that at least some compound chondrules could have been made by melting nebular dust
that had accreted onto a chondrule, which is a good model for enveloping \ccs, a
rare class of \ccs so named because the second chondrule mostly envelopes the first. 
\cite{2008Icar..194..811M} instead suggests a model where very large dust grains (far larger than observed chondrules)
had their surfaces melted and stripped by a shock in the nebular gas.  The stripped surface separated into
droplets which in turn collided with each other, creating compound chondrules.

Excepting the production of enveloping \ccs by melting an accretionary dust rim, the above models all
require that molten droplets remain in contact, 
with very small contact angles, for significant collisional time scales. Note that we use the term ``molten'' to refer to chondrules
sufficiently liquid that they collide as liquids, rather than solids.  This is important because it allows the individual chondrules
to collide at sufficiently elevated speeds that the observed \cc frequency might be matched.
In this paper, we consider only non-enveloping \ccs and show that even at temperatures well below chondrule formation temperatures
 \citep[$T \gtrsim 1700$\,K,][]{1990Metic..25..309H}, droplet coalescence time scales
are short compared to collisional time scales and that the molten-collision model cannot
match the observed \cc frequency. That means that the small contact angles observed are a major constraint, and imply
that \cc fusing had to occur relatively cold.

\section{Viscosity and surface tension}
Chondrules exhibit a range of sizes, compositions and textures \citep{1987JGR....92.E663W,1995GeCoA..59.1847W,Friedrich2014}.
Following previous studies
of \ccs \citep{2004M&PS...39..531C,2012M&PS...47.1139D}, we adopt a characteristic chondrule radius of $a=0.03$\,cm,
and a chondrule solid density $\rho=3$\gcm.

To understand the time scale on which molten chondrules in contact with each other would have flowed we need their surface tensions
and viscosities.
Chondrule melts had surface tensions on the order of
\EQ
\gamma=400 \text{\dcm} 
\EN
\cite[or $0.4$\,N\,m$^{-1}$,][]{2002ApJ...564L..57S}.
%
%
%


\begin{table}
\caption{H3 Chondrule Compositions \label{compositions}}
\centerline{\begin{tabular}{lD{.}{.}{-1}D{.}{.}{-1}}
\hline
 & \multicolumn{1}{c}{BO} & \multicolumn{1}{c}{Average} \\ 
\hline
SiO$_2$           & 45.4 & 51.8 \\ 
TiO$_2$           & 0.21 & 0.14 \\  
Al$_2$O$_3$  & 5.7   & 3.4   \\  
Cr$_2$O$_3$ & 0.41 & 0.55 \\ 
FeO                   & 14.0 & 9.1 \\    
MnO                  & 0.41 & 0.34   \\
MgO                  & 28.7 & 30.7 \\  
CaO                  & 2.24 & 2.30 \\  
Na$_2$O         & 2.43 & 1.40 \\  
K$_2$O              & 0.45 & 0.17 \\
\hline 
\end{tabular}}
\end{table}

\begin{figure}[t!]\begin{center}
\includegraphics[width=\columnwidth]{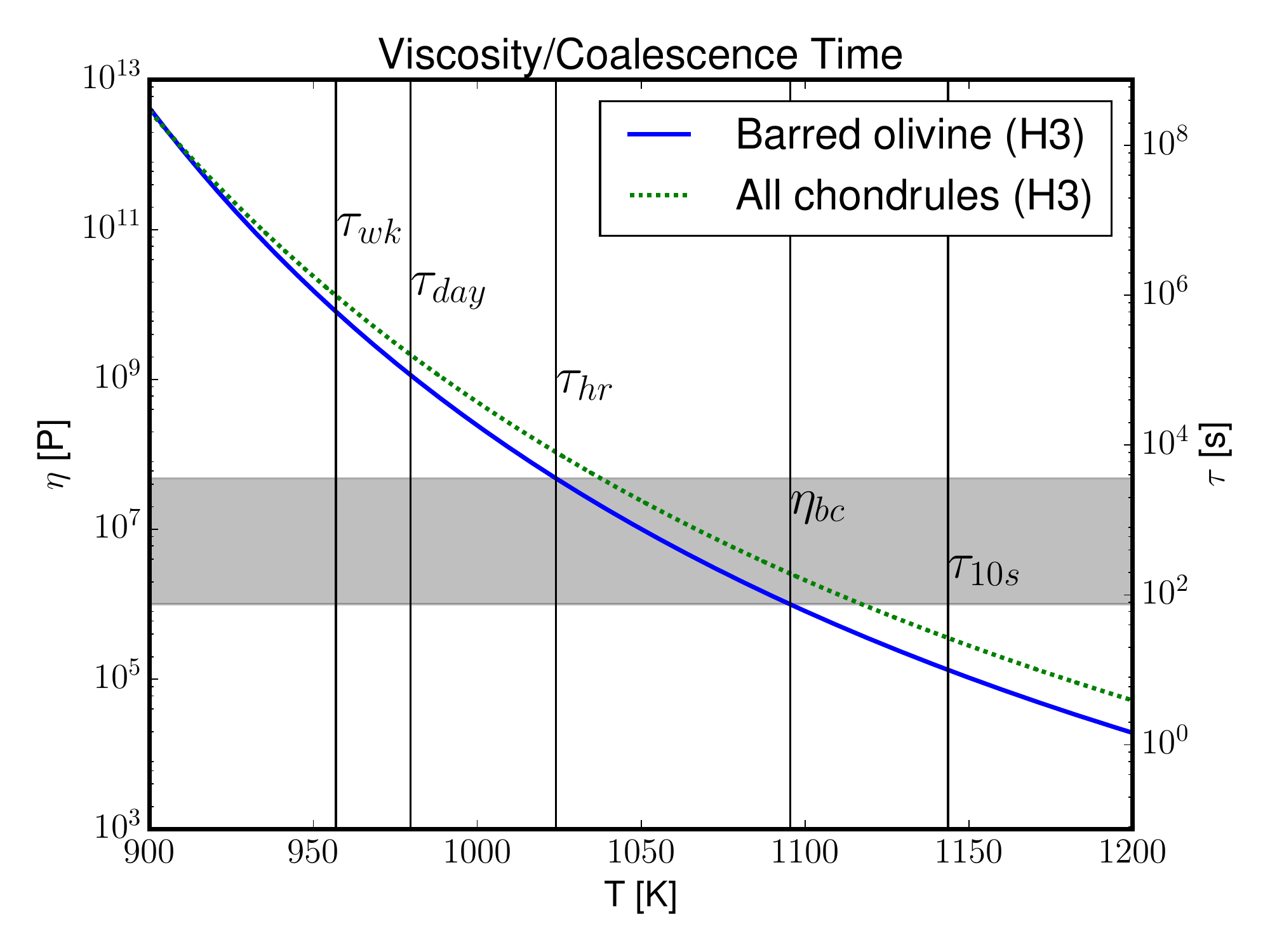}
\end{center}
\caption{
Viscosities (left axis) and coalescence times (right axis) for droplets with radii $a=0.03$\,cm.  Solid/blue: 
H3 barred olivine chondrule composition.  Dashed/green: H3 average chondrule composition.
Vertical lines denote the temperature below which colliding BO chondrules bounce, and the temperatures
at which BO chondrules have coalescence times $\tau$ of ten seconds, an hour, a day and a week.
The shaded area separates the viscosities below which
droplets are sufficiently molten to stick well from the viscosities above which the individual
droplets are sufficiently solid to maintain their structural integrity for meaningful time scales.
\label{fig_viscosities} 
}
\end{figure}

We use 
\cite{2008E&PSL.271..123G} to calculate viscosities of chondrule melts\footnote{At the time of writing, a convenient on-line calculator can be found at
http://www.eos.ubc.ca/~krussell/VISCOSITY/grdViscosity.html}. 
\cite{1987JGR....92.E663W} lists the composition of chondrules found in ordinary chondrites (see \Tab{compositions}).
We find
that the viscosities for barred olivine/average chondrules in ordinary chondrites were bracketed by the values of those
chondrules in H3 chondrites:
\EQ
\log \eta = -3.55 +\frac{4557.8}{T-618.2} \label{eta_BO}
\EN
and
\EQ
\log \eta = -3.55 +\frac{5084.9}{T-584.9} \label{eta_all}
\EN
for barred olivines and average chondrules, respectively.  Barred olivines were less viscous for relevant temperature ranges
(see \Fig{fig_viscosities}, left axis).
$T$ is the temperature in Kelvin and we are measuring the viscosity $\eta$ in poise ($1$\,P\,$=0.1$\,Pa$\cdot$s).  While barred olivines are
rare in the overall chondrule population \citep{1981Metic..16...17G}, using barred olivine
compositions is not as strange as it might seem: \cite{1995GeCoA..59.1847W} found that barred olivines are overrepresented
in compound chondrules by a factor of many.

There is clear evidence that chondrules experienced multiple
heating events \citep{1997M&PS...32..753J,2003GeCoA..67.2239W,2005ASPC..341..251J}, and the existence of relict grains
implies that at the temperatures we consider, many chondrules
would have had more liquid, lower viscosity regions mixed with more solid, higher viscosity ones.  However, as noted by \cite{2006M&PS...41.1347C},
it is the lowest viscosity that controls the viscous behavior of the grains with the fluids flowing around the more solid inclusions.


\cite{2004M&PS...39..531C} found
that chondrules behaved as liquids while colliding if 
their interaction time was longer than the Maxwell time
\EQ
\tau_M = \frac{\eta}{\mu},
\EN
where $\mu \simeq 4 \times 10^{10}$\,Pa is the shear modulus.
They estimated the interaction time using Herztian contact theory, finding
\EQ
t_{coll} = 2.87 \left( \frac{m_1m_2}{m_1+m_2}\right)^{\frac 25}
\left(\frac{1-\nu_1^2}{E_1}+\frac{1-\nu_2^2}{E_2}\right)^{\frac 25}
\left( \frac{a_1+a_2}{a_1a_2}\right)^{\frac 15}
v_c^{-\frac 15},
\EN
where $m \simeq 3.4 \times 10^{-4}$\,g, $\nu \simeq 0.25$, and $E \simeq 10^{11}$\,Pa are the chondrules' masses,
Poisson ratios and Young's moduli respectively.
For those values, and a collision velocity $v_c=100$\cms, we can calculate
\EQ
t_{coll} \simeq 3 \times 10^{-6}\,\text{s},
\EN
so the condition for liquid collision reduces to
\EQ
\eta \lesssim  \eta_{bc} =10^6\,\text{P}.
\EN
That viscosity constraint implies 
\EQ
T>1095\,K,
\EN
which value approximately applies for all collisional velocites
as $\eta$ is a very steep function of temperature and the
critical $\eta_{bc}$ value is a very weak function of $v_c$.  
Below that viscosity limit, or equivalently above that temperature limit, molten chondrules have a high
sticking probability even at velocities above $100$\cms. 
Below that temperature, chondrules collided as solids, bouncing unless the collisional
speed was below $0.1$\cms \citep{2010A&A...513A..56G}.

\section{Coalescence time scales}

Liquid droplets in contact with each other tend to coalesce into a single droplet.
This makes it problematic that liquid-sticking requires $\eta< \eta_{bc} =10^6$\,P: we need the molten chondrules to have behaved as liquids
for collisional interaction times, but not have behaved as liquids for far longer collisional or cooling windows.

As might be expected for a phenomenon with significant consequences for industry, there is a rich literature on
droplet coalescence spreading 
\citep{Frenkel1945,1979JPhD...12.1473T,1999JFM...401..293E,2005PhRvE..71a6309Y,2005PhRvL..95p4503A,2013PhFl...25a3102E}.
Viscous droplet coalescence or spreading
time scales depend on a coalescence time scale:
\EQ
\tau = \frac{a \eta}{\gamma},
\EN
where $a$ is the droplet radius (see \Fig{fig_viscosities}, right axis). This scaling can be understood on dimensional grounds:
larger and more viscous droplets take longer to coalesce, while the surface tension provides the energy source, and $\tau$ is
the combination of $a$, $\eta$ and $\gamma$ with the correct dimensions.  One can also
compare the energy dissipated by Stokes drag $F_S$ on a sphere of radius $a$ moving a distance $a$ at a speed $v$ through a medium of
viscosity $\eta$ to the surface energy of a droplet
with surface area $\sim 4 \pi a^2$ and surface tension $\gamma$:
\EQ
a F_S  = a \times 6 \pi \eta a v \sim 4 \pi \gamma a^2,
\EN
and then estimate
\EQ
\tau \sim \frac{a}{v} \sim \frac{a \eta}{\gamma}.
\EN
Calculating $\tau$ for $\eta=\eta_{bc}$ and $a=0.03$\,cm, we find $\tau=75$\,s.

The precise time dependence of the contact angle (or neck width) between two droplets remains a topic of research, but it is clear that there are multiple
regimes where the radius of the interface between the droplets depends on $\tau^n$ for varying $n$, possibly with logarithmic
corrections.  However, it is nonetheless also clear from the literature
that small contact angles can only be preserved for time scales 
\EQ
t \ll \tau,
\EN
which places an upper limit on how long molten chondrules can collide for.
While this condition has the opposite implication as the plasticity time
$t_{\text{plas}}$ of previous work \citep{1981Metic..16...17G,2004M&PS...39..531C}, it enters
into analyses of \cc collision rates in the same fashion.

We should note that surface diffusion and sintering provides an alternate route to coalesence,
but one with a characteristic time that scales as $a^4$, significantly more strongly than fluid coalesence with $\tau \propto a$
 \citep{1998PhRvL..80.2634E}.  As such, it is expected to act relatively slowly on the scale
 of $a \simeq 0.03$\,cm chondrules and indeed, extrapolating from \cite{2003Icar..164..139P}'s
 data for SiO$_2$ spheres, surface diffusion acts far slower than liquid coalescence for the chondrules we
 are interested in.

\section{Collisional compound chondrule formation}

We can estimate the frequency of compound chondrules
that resulted from collisions between molten chondrules, assuming perfect sticking:
\EQ
f = n \sigma v_c t, \label{freq_eq}
\EN
where $f$ is the compound chondrule fraction, $n$ the number density of molten chondrules,
$\sigma$ the chondrule-chondrule collisional cross section,
$v_c$ a characteristic  collision speed and $t$ the time window over which collisions lead to \ccs fusing. 
Note that to avoid coalescence we require $t \ll \tau$.

For
$a=0.03$\,cm chondrules, 
\EQ
\sigma = 4 \pi \left(2 \times a\right)^2 \simeq 0.045\,\text{cm}^2.
\EN
At $R=2.5$\,AU \citep{2007ApJ...671..878D}, 
the Desch Minimum Mass
Solar Nebula model provides a high end estimate for the gas density model of the Solar Nebula of
$\rho_g =2.3 \times 10^{-9}$\gcm at the midplane. 
An equal mass density in $a=0.03$\,cm, $\rho=3$\gcm chondrules implies
\EQ
n \simeq 6.8 \times 10^{-8}\,\text{cm}^{-3}. \label{n_over}
\EN
Chondrule-sized dust grains are difficult to concentrate
(see \citealt{2012Icar..220..162J,2015Icar..245...32H}, for discussions of dust concentration focusing on chondrules),
so \Eq{n_over} is a safe, significant over-estimate for the chondrule number density.

Using disaggregated chondrules, \cite{1981Metic..16...17G} estimated that $\sim 4\%$ of chondrules are compound.
\cite{1995GeCoA..59.1847W} found a lower rate, $\sim 2.4\%$ by studying thin sections.  Subsequently \cite{2004M&PS...39..531C}
improved the 2D to 3D correction factors, finding instead $5\%$.  We split the difference and use $f=0.045$.
However, this frequency is not the entire story.  As noted by \cite{1981Metic..16...17G} and especially by
\cite{1995GeCoA..59.1847W}, compound chondrule do not appear to have been made by randomly
picking two candidates from the pool of all chondrules and sticking them together.  Radial pyroxene and barred olivine chondrules
are heavily over-represented: combined they make up 97 of the 144 non-enveloping chondrules listed in \cite{1995GeCoA..59.1847W}'s Table 7,
despite making up only $10-13\%$ of chondrules over all \citep{1981Metic..16...17G}.  If this result holds, it implies that those chondrule
classes have compound chondrule frequencies higher than the average by a factor of over $5$, and $f=0.045$ is
an underestimate of the required \cc frequency.

For viscosities below $\eta<\eta_{bc}$ (i.e.~temperatures above $T>1095$\,K), chondrules collide as liquids, and can stick
even for elevated collisional velocities.
However, at $T=1095$\,K, $\tau \simeq 75$\,s, while at only slightly
hotter temperatures, $T \simeq 1145$\,K,
$\tau=10$\,s.  We require $t \ll \tau$ to avoid
significant coalescence of the component chondrules into a single whole, and use $t=10$\,s as an overestimate,
because it can only apply to chondrules which collide just above $T=1095$\,K.
Chondrule precursors, loosely stuck together clumps of dust, could not survive
collision speeds above $v_c=100$\cms \citep{2010A&A...513A..56G}, constraining the background chondrule-chondrule
relative velocity.  However, several km\,s$^{-1}$ shocks, one of the proposed chondrule formation
processes \citep{1991Icar...93..259H,1998M&PS...33...97H}, can lead to significantly higher collision speeds \citep{2004LPI....35.1847N,2006M&PS...41.1347C}.  Shocks capable of reaching
the temperatures required for chondrule formation however also keep the chondrules above $1145$\,K for hundreds of seconds to tens of minutes
or even longer
\citep{2013ApJ...776..101B,2014Icar..242....1S}.  This is many times the chondrule stopping time in the gas of under a minute, so in a shock-melting
scenario, once the temperature had fallen below $1150$\,K, the chondrule relative velocity would have fallen far below
the shock's intrinsic velocity of several km\,s$^{-1}$.  We therefore use $v_c=10^4$\cms as an upper limit to chondrule relative velocites;
while noting that restricting ourselves to temperatures in the range of $1095-1145$\,K and time scales of $t=10$\,s ignores the far longer time scales
spent at higher temperatures, which would have yielded significantly more collisions.  Those collisions would often have been destructive \citep{2014ApJ...797...30J},
and even when they led to sticking, would have resulted in coalescence, rather than \cc fusing.

Solving \Eq{freq_eq} for $f$, using $t=10$\,s, $v_c=10^4$\cms
and the above values for $n$ and $\sigma$, we find
\EQ
f = 0.03.
\EN
While this is close to the observed value, it requires a significant overestimate in the chondrule number density.  Further,
it requires that all the collisions occur at temperatures above $T=1095$\,K but significantly below $T=1145$\,K while cooling out
of that window on times neither much longer nor shorter than $10$\,s.  
Many chondrules experienced multiple
heating events \citep{1997M&PS...32..753J,2003GeCoA..67.2239W,2005ASPC..341..251J},
but any event which remelted a chondrule ($T \gtrsim 1700$\,K)
would have also led to coalescence, leaving only the \ccs that fused following the final heating event.
Accordingly, we can safely discount molten-chondrule collisions
as a \cc formation mechanism.

For viscosities above $\eta>\eta_{bc}$ (i.e.~temperatures below $T<1095$\,K), chondrules collide as solids, and will only stick for collision velocities
\citep{2010A&A...513A..56G}
\EQ
v_c \lesssim 0.1\,\text{\cms.}
\EN
Using \Eq{freq_eq} to match $f=0.045$ then requires
\EQ
t \simeq 1.5 \times 10^6\,\text{s,}
\EN
or over two weeks.  The requirement that $t \ll \tau$ then implies $T<950$\,K.  Note further that we have overestimated $n$,
and that the overall disk temperature at $R=2.5$\,AU was much colder than $950$\,K \citep{1981PThPS..70...35H,2007ApJ...671..878D}.
In the solid-collision scenario it is far easier for the component chondrules to have collided 
at temperatures far below $950$\,K (with almost arbitrarily long time scales available)
and been subsequently heated to temperatures
above $T>900$\,K for just long enough for the first hints on coalescence to have occurred.

\section{Conclusions}

\Fig{fig_viscosities} shows that for any reasonably warm temperature, there is a fusing time scale for chondrules
in contact to form non-enveloping \ccs.  If the potential \cc remained at that temperature for too short a time, then it would not have fused, while
if it remained at that temperature too long the individual components would have coalesced. The molten-collision model for \cc formation
makes testable predictions about the temperature of the chondrules upon collision (they have survived the collision and stuck together) and the time scales available
(sufficient \ccs must have been produced).
The shaded band in \Fig{fig_viscosities} shows that if ``molten'' chondrules are sufficiently inviscid that they can behave as liquids upon collision and stick ($\eta <\eta_{bc}$),
they are far too inviscid not to coalesce too fast for a sufficient number of compound chondrules to have been produced.
This constraint does not depend on the precise temperature
dependence of the viscosity $\eta$, only the critical viscosity $\eta_{bc}$ above which chondrules collide as solid, not liquid, spheres, and hence
is a robust result. 

If compound chondrules could not have resulted from the collisions of molten chondrules, and indeed the primary and secondary chondrules 
could not ever have been in contact at temperatures significantly above $\sim 1000$\,K, then they must be the result of chondrules
colliding and sticking while cold.  While naked chondrules are not expected to have been particularly sticky, experiments have shown that dusty
rims could have bypassed that limitation \citep{2012Icar..218..701B}.
Even so, sticking would have required low relative velocities, and
hence the collision rate would have been very low, compensated for by almost arbitrarily long time scales chondrules could have
spent floating freely and cold in the Solar Nebula.

In \Fig{fig_viscosities}, we have marked the temperatures for which the coalescence time
$\tau$ is ten seconds, one hour, day and week, as well as the temperature below which bouncing is expected.
The evidence that chondrules saw multiple heating events \citep{2005ASPC..341..251J} implies
that many such pairs of stuck together (but not yet fused together) chondrules would have experienced episodes
of elevated temperature.  If those episodes were a small but still respectable fraction of the coalescence time of their peak temperature, the component
chondrules would have fused, but not coalesced, leaving a \cc.
The temporal longevity and strength of temperature fluctuations in protoplanetary disks are not well understood, but
it seems unlikely that a proto-compound chondrule could be held at temperatures above $\sim 1025$\,K only for time scales short enough not
to coalesce (no more than an hour).
Similarly, it seems unlikely that a proto-\cc could be held at temperatures just above, but only just above, $900$\,K  for any significant fraction
of the corresponding $\tau=1$\,decade, so at colder temperatures there would be inadequate spreading to allow \ccs to fuse.
We also know from noble gas measurements that only a modest fraction of the solids that were
eventually incorporated into chondrites ever saw temperatures above $\sim 800$\,K \citep{2015Icar..245...32H}, further constraining the ability to store proto-\ccs
in regions with $T>800$\,K for prolonged periods. We conclude that compound chondrules fused in a temperature window 
of $900-1025$\,K, with very finite dwell times.

The different viscosities expected for chondrules of average composition and BO chondrules (due to the later's smaller
SiO$_2$ fraction) could help explain one of the mysteries associated with \ccs: the fraction of radial pyroxene
and barred olivine chondrules in \ccs appears to be
several times higher
than in the overall chondrule population \citep{1995GeCoA..59.1847W}.  If that result holds, the participant pairs in \ccs were not selected at random.
BO chondrules
are expected to have been less viscous than an average chondrule (\Fig{fig_viscosities}),
which means that they would have been more prone to spreading.
That would have allowed BO chondrules to partially fuse and form \ccs as a result of temperature fluctuations too weak to affect more viscous
chondrules.


The frequency of compound chondrules cannot be used
as a probe of the chondrule density in melting regions, as it has been used.  However, with better statistics
\ccs could potentially be used as a probe of cold
chondrule-chondrule sticking rates.  This is important as chondrule agglomerations are an significant part of chondrite formation, and
there are indications that chondrules did not dwell for long in chondrule forming regions \citep{2015Icar..245...32H} after forming.
Similarly, this means that \ccs provide a potential probe for a new temperature window in the Solar nebula:
noble gases in trapped in chondrite matrix \citep{1994Metic..29..811H} and primary troilite in chondrules 
\citep{1999GeCoA..63.2281R} place
upper experienced-temperature limits of about $800$ and $650$\,K respectively, significantly colder than our $900-1025$\,K.  This new temperature
window is especially interesting as it lies near the $\sim 1000$\,K threshold for thermal ionization to
allow the magneto-rotational instability to operate
\citep{1996ApJ...457..355G}. These temperatures can also be further constrained by considering isotopic fractionation or lack
thereof \citep{2000M&PS...35..859A} because our temperature window is close to the $50\%$ condensation
temperature of potassium in the Solar Nebula \citep[$\sim 1006$\,K,][]{2003ApJ...591.1220L}.

\section*{Acknowledgments}

I thank Michael Weisberg and Denton S. Ebel for \Fig{fig_pic}.  The two referees both provided feedback which greatly enhanced the
clarity of the arguments.
This research has made use of the National Aeronautics and Space AdministrationÕs Astrophysics Data System Bibliographic Services.
The work was supported by National Science Foundation, Cyberenabled Discovery Initiative grant AST08-35734, 
AAG grant AST10-09802, NASA OSS grant NNX14AJ56G and a Kalbfleisch Fellowship from the American Museum of Natural History.

\bibliography{paper}

\end{document}